\documentclass[a4paper]{jpconf}
\usepackage{graphicx}

\newcommand{\lwig}{\mbox{\,\raisebox{.3ex}
    {$<$}$\!\!\!\!\!$\raisebox{-.9ex}{$\sim$}\,}}

\newcommand{\lambdabar}{{\hbox{$\lambda_e$\kern-1.9ex\raise+0.45ex\hbox{--}
\kern+0.2ex}}}

\begin{document}
\title{{\vskip -1cm\normalsize\rm\hfill DESY 05-229} 
\vskip 1cm 
Axion interpretation of the PVLAS data?\footnote{Talk presented at the ninth 
International Conference on Topics in Astroparticle and Underground Physics, TAUP 2005, Zaragoza, Spain, September 10-14, 2005.}}

\author{Andreas Ringwald}

\address{Deutsches Elektronen-Synchrotron DESY, Notkestra\ss e 85, D--22607 Hamburg, Germany}

\ead{andreas.ringwald@desy.de}

\begin{abstract}
The PVLAS collaboration has recently reported the observation of a
rotation of the polarization plane of light propagating through a
transverse static magnetic field. Such an effect can arise from the
production of a light, $m_A\sim$~meV, pseudoscalar coupled to two photons
with coupling strength $g_{A\gamma}\sim 5\times 10^{-6}$~GeV$^{-1}$. 
Here, we review these experimental findings, discuss how astrophysical 
and helioscope bounds on this coupling can be evaded, and 
emphasize some experimental proposals to test the scenario.  
\end{abstract}

There are various proposals in the literature in favour of the existence
of light pseudosca\-lar particles beyond the Standard 
Model which have, so far, remained undetected, due to their weak coupling 
to ordinary matter. Such light particles would arise if there was a global continuous 
symmetry in the theory that is spontaneously broken in the vacuum. 
A well known example is the axion~\cite{Weinberg:1978ma}, which 
arises from a natural solution to the strong $CP$ problem.
It appears as a pseudo Nambu-Goldstone boson of a spontaneously broken Peccei-Quinn
symmetry~\cite{Peccei:1977hh}, whose scale $f_A$ determines its mass,
${m_A} = [z^{1/2}/(1+z)]\,
         m_\pi f_\pi/ f_A= { 0.6\,  {\rm meV}} 
         \times
         ( 
         10^{10}\, {\rm GeV}/{ f_A}  
        )
$
in terms of the mass $m_\pi$ and decay constant $f_\pi$ of the pion
and the current quark mass ratio $z=m_u/m_d$. 
Only invisible axion 
models~\cite{Kim:1979if,Zhitnitsky:1980tq}, 
where $f_A\gg 247$ GeV, are viable experimentally~\cite{Eidelman:2004wy}.

\begin{figure}[t]
\begin{center}
\includegraphics*[bbllx=25,bblly=226,bburx=564,bbury=604,width=15.5cm]{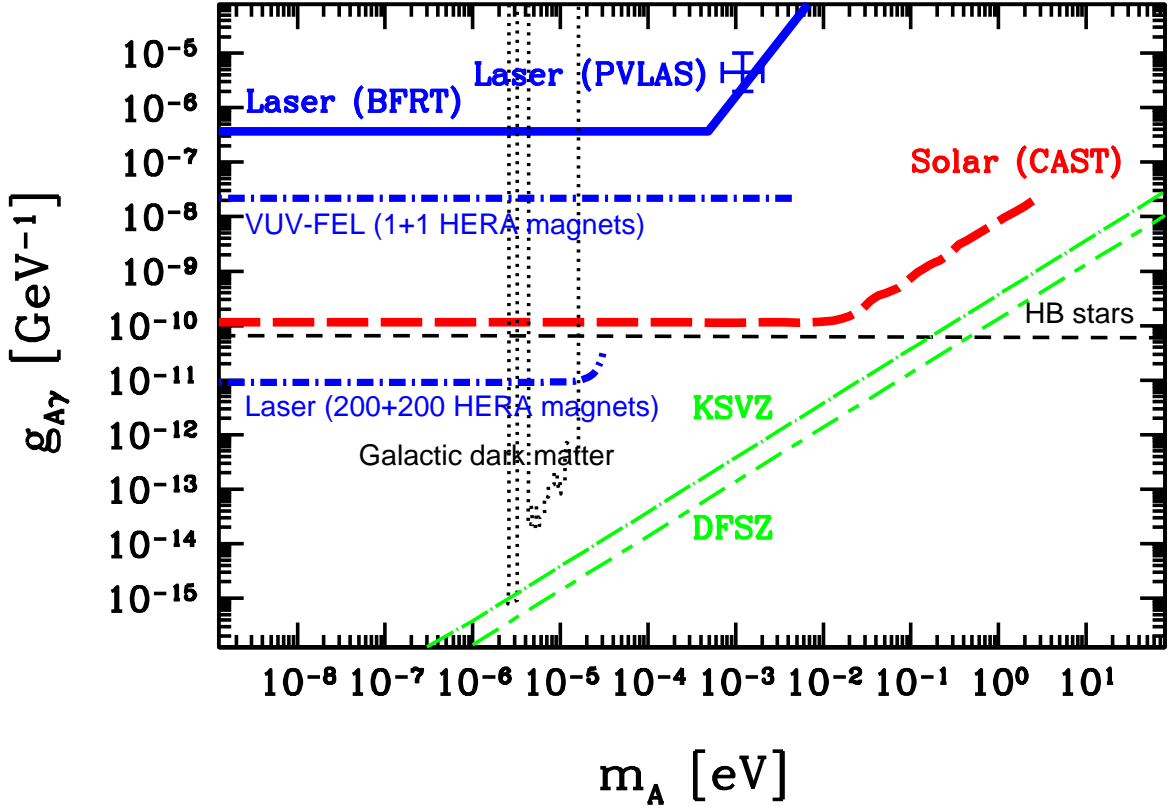}
%\vspace{-.45cm}
\caption[...]{Exclusion region in mass $m_A$ vs. axion-photon coupling $g_{A\gamma}$ 
for various current and future experiments.
The laser experiments~\cite{Cameron:mr,Zavattini:2005tm,Ringwald:2001cp,Ringwald:2003ns} aim at  
axion production and detection in the laboratory.  
The galactic dark matter experiments~\cite{Eidelman:2004wy} exploit microwave cavities to detect axions 
under the assumption that axions are the dominant constituents of our galactic halo, 
and the solar experiments 
search for axions from the sun~\cite{Andriamonje:2004hi}. 
The constraint from horizontal branch (HB) stars~\cite{Eidelman:2004wy,Raffelt:1999tx} 
arises from a consideration of stellar  
energy losses through axion production.
\hfill
\label{fig:ax_ph}}
\end{center}
\end{figure}

Clearly, it is of great interest to set stringent constraints on the properties
of such light pseudoscalars.  
The interactions of axions and similar light pseudoscalars with Standard Model particles are model 
dependent, i.e. not a function of $1/f_A$ only.  
The most stringent constraints to date come from their coupling to photons, 
$g_{A\gamma}$, which arises via the axial anomaly~\cite{Bardeen:1977bd}, 
\begin{equation}
\label{eq:ax_ph}
        {\mathcal L}_{\rm int} = 
-\frac{1}{4}\,{ g_{A\gamma}}\,A\ F_{\mu\nu} \tilde{F}^{\mu\nu} 
=
 { g_{A\gamma}}\,A\ {\mathbf E}\cdot {\mathbf B}\, ;
        \hspace{5ex}
        { g_{A\gamma}} = -\frac{\alpha}{2\pi { f_A}} 
\left( { \frac{E}{N}} - \frac{2}{3}\,\frac{4+z}{1+z}\right)
\,,
\end{equation}
where $A$ is the pseudoscalar field, $F_{\mu\nu}$ ($\tilde{F}^{\mu\nu}$) the (dual) electromagnetic field strength tensor, 
$\alpha$ the fine-structure constant, and 
$E/N$ the ratio of electromagnetic over color anomalies. 
As illustrated in Fig.~\ref{fig:ax_ph}, 
two quite distinct invisible axion models, namely the KSVZ~\cite{Kim:1979if} 
(or hadronic) and the DFSZ~\cite{Zhitnitsky:1980tq} (or grand unified) one, 
lead to quite similar $g_{A\gamma}$.    
The strongest constraints currently involve cosmological and astrophysical 
considerations.  Only the laser experiments in Fig.~\ref{fig:ax_ph} aim also at the production 
of axions in the laboratory.  

%%%%%%%%%figure%%%%%%%%%%%%%%%%
%%%%%%%%%%%%%%%%%%%%%%%%%%%%

Let us discuss such laser experiments in some detail. 
The most straightforward ones exploit photon regeneration.  
They are  based on the 
idea~\cite{Sikivie:ip}
to send a polarized laser beam, with average power $\langle P\rangle$ and frequency $\omega$, 
along a superconducting dipole magnet of length $\ell$,  
such that the laser polarization is parallel to the magnetic field. 
In the latter, the photons may convert into axions via 
a Primakoff process.  
If another identical dipole magnet is set up in line with the first magnet, with a sufficiently thick
wall between them to absorb the incident laser photons, 
then photons may be regenerated from the pure axion beam
in the second magnet 
and detected with an efficiency $\epsilon$.  
The expected counting rate of such an experiment is given by 
\begin{equation}
\label{eq:ax_counting_rate}
\frac{{\rm d}N_\gamma}{{\rm d}t} =
{\frac{\langle P\rangle }{\omega}}\ 
\frac{N_r+2}{2}\,
\frac{1}{16} \left( g_{A\gamma}\,{B}\,\ell\right)^4 
\sin^2
\left( \frac{m_A^2\,\ell }{4\,\omega }
\right) 
\left( \frac{m_A^2\,\ell }{4\,\omega }
\right)^2  
\approx  { \frac{\langle P\rangle }{\omega}}\ 
\frac{N_r+2}{2}\,
\frac{1}{16} \left( g_{A\gamma}\,{B}\,\ell\right)^4 
\eta         
\,,
\end{equation}
if one makes use of the possibility of putting the first magnet into an 
optical cavity with a total number $N_r$ of reflections.
For 
$m_A \ll \sqrt{2\,\pi\,\omega/\ell} = 4\times 10^{-4}\, {\rm eV}
    \sqrt{({ \omega}/1\, {\rm eV})  (10\, {\rm m}/\ell )  }$, 
the approximate sign in (\ref{eq:ax_counting_rate}) applies and  
the expected counting rate for a photon regeneration experiment 
is independent of the axion mass. 
A pilot photon regeneration experiment was performed by the 
Brookhaven-Fermilab-Rutherford-Trieste (BFRT) collaboration~\cite{Cameron:mr}.  
It employed an optical laser
of wavelength $\lambda =2\pi/\omega = 514$~nm and power $\langle P\rangle = 3$~W for 
$t=220$~minutes in an optical 
cavity with $N_r=200$, and used two superconducting 
dipole magnets with $B = 3.7$ T and $\ell = 4.4$ m. 
No signal of photon regeneration was found, which leads, taking into account a 
detection efficiency of 
$\eta =0.055$, to a $2\,\sigma$ upper limit of 
$g_{A\gamma}<6.7\times 10^{-7}$~GeV$^{-1}$ for axion-like pseudoscalars with
mass $m_A< 10^{-3}$ eV. 

Another possibility to probe $g_{a\gamma}$ is to 
measure changes in the polarization state when photons have traversed 
a transverse magnetic field~\cite{Maiani:1986md}.  In particular, the 
real production of axions leads to a rotation of the polarization plane of an 
initially linearly polarized laser beam by an angle 
\begin{eqnarray}
\label{ax_rot}
\epsilon &=& N_r\,\frac{g_{A\gamma}^2\,B^2\,\omega^2}{m_A^4}\,
\sin^2\left( \frac{m_A^2\,\ell}{4\,\omega}\right)\,
\sin 2\,\theta
\approx \frac{N_r}{16} \left( g_{A\gamma}\,B\,\ell \right)^2\,\sin 2\,\theta 
\,,
\end{eqnarray}   
where $\theta$ is the angle between the light polarization direction and the 
magnetic field component normal to the light propagation vector. 
The BFRT collaboration has also performed a pilot polarization experiment along these lines, 
with the same laser and magnets described before. For $\ell = 8.8$~m, $B=2$~T, and $N_r=254$, 
an upper limit on the rotation angle $\epsilon< 3.5\times 10^{-10}$~rad 
was set, leading to a 
limit $g_{A\gamma}< 3.6\times 10^{-7}$~GeV$^{-1}$ at the 95\,\% confidence level, 
provided $m_A<1$~meV~\cite{Cameron:mr}. Similar limits have been set from the absence of 
ellipticity in the transmitted beam. The overall envelope of the constraints from 
the BFRT collaboration~\cite{Cameron:mr} is shown
in Fig.~\ref{fig:ax_ph} and labelled by ``Laser (BFRT)'' (cf. Ref.~\cite{Eidelman:2004wy}). 

Recently, the PVLAS experiment~\cite{Zavattini:2005tm}, consisting of a 
Fabry-P\'erot cavity of very high finesse ($N_r\approx 44\, 000$),  immersed in a 
magnetic dipole with $\ell =1$~m and $B=5$~T, reported 
the observation of a
rotation of the polarization plane of light propagating through a
transverse static magnetic field~\cite{Zavattini:2005tm}. If interpreted in terms of the production 
of a light neutral pseudoscalar, the PVLAS collaboration finds a region
$1.7\times 10^{-6}\, {\rm GeV}^{-1}\,\lwig\,g_{A\gamma}\,\lwig\,1.0\times 10^{-5}\, {\rm GeV}^{-1}$ 
for $0.7\, \textrm{meV}\,\lwig\, m_A\,\lwig\, 2.0\, \textrm{meV}$,  from a combination of
the $g_{A\gamma}$ vs. $m_A$ curve corresponding to the PVLAS rotation signal (cf. Eq.~(\ref{ax_rot})) 
with the BFRT limits on the same quantities. 

Clearly, a pseudoscalar with these properties is hardly compatible with a genuine QCD axion. 
For the latter, a mass $m_A\sim 1$\,meV implies a symmetry breaking scale 
$f_A\sim 6\times 10^{9}$\,GeV. According to (\ref{eq:ax_ph}), one needs then an extremely large ratio 
$|E/N|\sim 3\times 10^7$ of electromagnetic and color anomalies in order to arrive at an axion-photon 
coupling in the range suggested by PVLAS.
This is far away from the predictions of any model conceived so far~\cite{Cheng:1995fd}. 
Moreover, such a pseudoscalar must have very peculiar properties in order to evade the 
strong constraints on $g_{A\gamma}$ from stellar energy loss considerations 
(``HB stars" in Fig.~\ref{fig:ax_ph}) and from its non-observation 
in helioscopes such as the CERN Axion Solar Telescope 
(``Solar (CAST)" in Fig.~\ref{fig:ax_ph})~\cite{Raffelt:2005mt}. 
Pseudoscalar production in stars may be hindered, for example, 
if the $A\gamma\gamma$ vertex is suppressed at keV energies due to low scale 
compositeness of $A$~\cite{Masso:2005ym} or if, in stellar interiors, 
$A$ acquires an effective mass larger than the typical photon energy, $\sim$~keV~\cite{jaeckel:tbp}.
   
In any case, an independent and decisive experimental test of the 
finding of PVLAS is urgently needed. One opportunity is offered by high luminosity $e^+e^-$ colliders,  
e.g. a possible super-$B$ factory at KEK, where one may search  
for  events with a single photon plus missing transverse energy in
the final state~\cite{Masso:1997ru}. 
The best and most timely possibilities, however, are offered by dedicated photon regeneration experiments, 
either based on ordinary optical lasers (e.g.~\cite{Duvillaret:2005sv}) or on 
(soft) X-rays from free-electron lasers (FEL) at DESY and SLAC~\cite{Ringwald:2001cp}.     
In fact, as can be seen in Fig.~\ref{fig:ax_ph}, the region of parameter space implied by PVLAS could be probed in a matter of 
minutes if one sets up a photon regeneration experiment exploiting the already operating FEL at 
DESY's TESLA Test Facility, which provides tunable radiation from the vacuum-ultraviolet (VUV) 
to soft X-rays,  $\omega=10$--$200$~eV, with an average power $\langle P\rangle = 20$--$40$~W,  
together with two superconducting dipole magnets of the type used in DESY's electron proton 
collider HERA ($B=5$~T, $\ell =10$~m)~\cite{Ringwald:2001cp}. 
The tuning of the FEL for  fixed photon flux would allow a precision determination of $m_A$. 
Such an experiment 
could also serve as a test facility for an ambitious large scale photon
regeneration experiment with sensitivity exceeding CAST~\cite{Ringwald:2003ns}, based on the recycling of all the 400 dipole magnets of HERA after 
its decommissioning in mid of 2007.

\end{document}